\documentclass[aps,pra,10pt,superscriptaddress,twocolumn]{revtex4-2}
\usepackage{graphicx}
\usepackage{amsmath,amsthm,amssymb,dsfont}
\usepackage{mdframed}
\usepackage{orcidlink}

\usepackage{hyperref}

\begin{document}

\title{Gravity-induced entanglement under constrained dynamics}

\author{Hollis Williams \orcidlink{0000-0003-3292-602X}}
\email{holliswilliams@hotmail.co.uk}

\affiliation{Theoretical Sciences Visiting Program, Okinawa Institute of Science and Technology Graduate University, Onna-son, Okinawa 904-0495, Japan}

\affiliation{Department of Mathematics and Statistics, University of Exeter, Exeter EX4 4QF, UK}

\affiliation{School of Engineering, Westlake University, Hangzhou, Zhejiang 310030, China}

\begin{abstract}
Tests of gravity-induced entanglement have been proposed as a route to probing the quantum nature of gravity, but existing schemes rely on free-fall interferometry of massive spatial superpositions, imposing severe experimental constraints. We show that systems exhibiting effectively inertial dynamics in the short-time regime reproduce the same gravitational phase accumulation responsible for entanglement generation. Deviations from the free-fall phase enter at order $(\tau/T)^2$, where $\tau$ is the interaction timescale and $T$ is the characteristic period of the constrained motion. We analyse a representative mechanically constrained implementation using carbon nanotube pendula and show that the resulting correction to the entangling phase remains small in experimentally relevant regimes, leading to a negligible modification of the interference visibility used to certify entanglement. These results demonstrate that gravity-induced entanglement protocols extend beyond free-fall implementations to a broader class of constrained dynamical systems, complementing existing proposals for experimental realisations of the Bose-Marletto-Vedral protocol.
\end{abstract}
\maketitle

\textit{Introduction} - Two recent proposals have outlined experimental tests of the quantum nature of gravity by investigating whether gravitational interactions can generate entanglement between a pair of massive particles \cite{bose, vedral, carney}.  In these schemes, each particle is prepared in a spin superposition and passes through a Stern-Gerlach interferometer, resulting in a massive spatial superposition.  The relative positions of the particles differ across the branches of the superposition, leading to branch-dependent gravitational interactions. If gravitational fields can exist in a quantum superposition, these interactions give rise to phase differences which entangle the spin degrees of freedom during recombination.  Crucially, it is a fundamental result in quantum information theory that entanglement between spatially separated systems cannot be generated by local operations and classical communication (LOCC) alone \cite{nielsen, horodecki}.  It follows that the observation of entanglement in such experiments would imply that gravity must act as a non-classical mediator, providing evidence that it possesses quantum features \cite{nguyen, chevalier}.

This protocol has been extended in many directions and is an active topic of research \cite{dekamp, schut1, schut2, ghosal, liu}.  However, explaining how to implement the concept even in a semi-realistic experimental setup is a problem which has so far proved prohibitively difficult to solve.  One of the main problems is how to practically create the massive spatial superposition using a Stern-Gerlach interferometer  \cite{machluf, margalit1, margalit2}.  Scala et al. introduced the idea of using an optically levitated diamond bead with a single nitrogen-vacancy (NV) center spin to create this superposition \cite{scala}.  This was developed into a free flight scheme based on coherent spin control which can be used to obtain a macroscopic spatial superposition \cite{wan}.  

Pedernales et al. later observed that the diamagnetic oscillations induced in the diamond by the Stern-Gerlach interferometer severely restrict the amount of time for which the particle can be in a spatial superposition in this scheme \cite{pedernales}.  As a possible resolution, it was suggested to incorporate spin dynamical decoupling \cite{wood}.  This setup requires a complicated and very specific arrangement of magnetic teeth to create an inhomogeneous field for part of the drop and also requires a large drop distance over which fine experimental control of a massive superposition would be extremely unrealistic (one second of free fall requires a 5 m drop and on the length scale of meters, temperature fluctuations would alter the length of the tower on a scale well above the size of the diamond).

In this article, we introduce a mechanically constrained model for gravity-induced entanglement, motivated by a pendulum setup (shown in Fig. 1). Concretely, one may consider two micron-scale diamonds containing single NV centers, each attached to the end of a thin carbon nanotube acting as a simple pendulum. Each diamond is also combined with a small paramagnetic compensator. Given that the mass of the composite particle is much larger than that of the nanotube, the system may be accurately modelled as a simple pendulum with the mass effectively concentrated at its end.  Nanotubes of length $0.5$ m and diameter on the order of $ 1\,\mathrm{nm}$ have been realised experimentally, so this regime is feasible \cite{zhang}.

In contrast to free-fall implementations, the present scheme enables repeated operation under fixed and well-controlled conditions, allowing long integration times and precise calibration. Moreover, the suppression of diamagnetic oscillations in the compensated system removes a key limitation on superposition size identified in free fall proposals \cite{pedernales}.  Levitation and trapping of microscale particles is also no longer required \cite{gonzalez, lialys}.

\begin{figure}
  \centering
   \includegraphics[width=80mm]{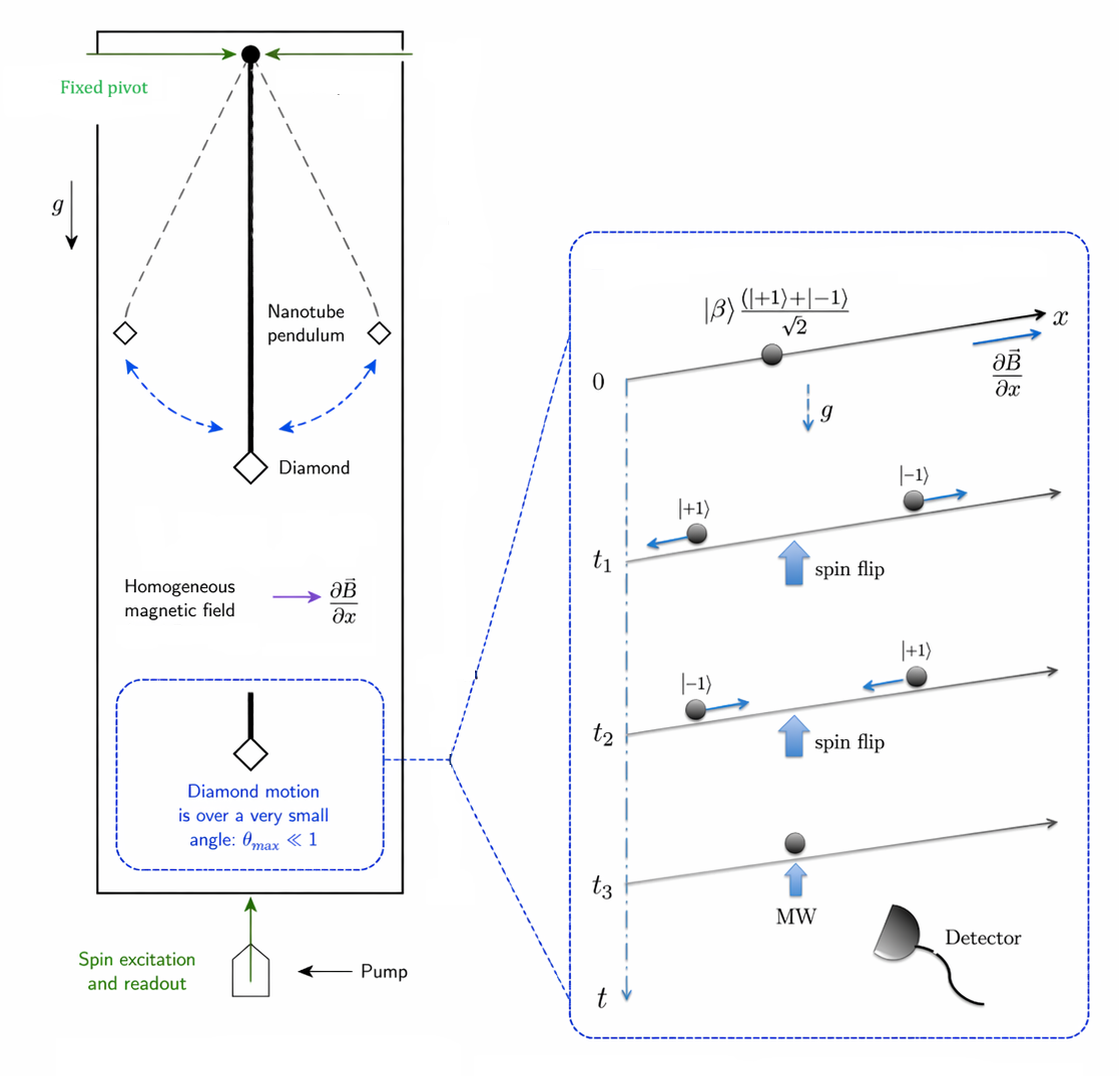}

 \caption{
 Schematic of the proposed nanotube-based platform. (Left) A microdiamond containing an NV center is attached to the end of a long carbon nanotube, forming a pendulum with motion confined to a small angular range ($\theta_{\mathrm{max}} \ll 1$).  Over the duration of the interferometric sequence, the constrained motion can be approximated by its short-time, locally inertial evolution. A magnetic-field gradient couples the spin state of the NV center to the center-of-mass motion. (Right) Schematic of the corresponding spin-dependent interferometric sequence, showing spatial splitting, spin flips and recombination. The sequence is illustrated using the free-flight Stern–Gerlach protocol of Wan et al. as a representative preparation scheme \cite{wan}.  The present work does not assume the numerical parameters of that proposal and recent proposals demonstrate that rapid generation of macroscopic spatial superpositions over a broader range of masses and micrometre-scale separations is feasible \cite{xiang}. Inset adapted from C. Wan et al., Phys. Rev. Lett. 117, 143003 (2016), DOI: 10.1103/PhysRevLett.117.143003, licensed under CC BY 3.0.
}

\end{figure}

The proposal we suggest here is distinct from other works which deal with nanofabricated torsion pendulums to detect quantum gravity effects or experiments which consider effects of minimum length scales (such as those given by candidate theories of quantum gravity) on the period of a pendulum in magnetic field gradients \cite{kumar, manley}.  In the proposal of \cite{kumar}, the mass at the end of the pendulum is also a composite particle formed of a nanodiamond with an embedded NV center combined with a paramagnetic material to allow magnetic compensation, but the type of experiment they propose is fundamentally different to ours, which is based on witnessing whether gravity can entangle two such pendulum-mounted composite particles.  It is also distinct from recent work exploring alternative attempts to probe the nature of gravity by identifying observable signatures which distinguish classical from quantum descriptions of the gravitational interaction without relying on entanglement generation as a witness \cite{kryhin}.  More generally, the construction in this article should be viewed as an effective model of mechanically constrained entanglement, with the nanotube realisation providing a concrete example which could be experimentally feasible.

\textit{Short-time constrained dynamics} - We consider a mechanically constrained system where the spin-dependent center-of-mass motion is confined by a pendulum, but approaches locally inertial motion over timescales which are short compared with the period of the pendulum $T$.  For a pendulum of length $L$, the small-angle angular coordinate $\theta$ of the pendulum satisfies

\begin{equation}
    \ddot{\theta}+\omega^2 \theta=0,
\end{equation}

\noindent
where $\omega=2\pi/T=\sqrt{g/L}$.  Given an initial angular displacement $\theta_0$ and vanishing initial angular velocity $\dot{\theta}_0$, the solution is

\begin{equation}
    \theta(t)=\theta_0\cos(\omega t).
\end{equation}

\noindent
Expanding for short times gives

\begin{equation}
    \theta(t)
    =
    \theta_0
    \left(
    1-\frac{\omega^2t^2}{2}
    +\frac{\omega^4t^4}{24}
    +\cdots
    \right).
\end{equation}

For a pendulum with length $L$, the corresponding arc displacement measured along the interferometeric axis may be written as $s = L \theta$, hence

\begin{equation}
    s(t)
    =
   s_0
    \left(
    1-\frac{g}{2L}t^2
    +\frac{g^2}{24L^2}t^4
    +\cdots
    \right).
\end{equation}

\noindent
Up to quadratic order, we therefore have

\begin{equation}
   s(t) \simeq s_0 \bigg( 1   -\frac{g}{2L}t^2 \bigg).
\end{equation}

\noindent
In this work, we compare the constrained spin-dependent center-of-mass displacement and the corresponding locally inertial trajectory with the same initial conditions, where the leading deviation between the trajectories arises at higher order in the short-time expansion (the resulting correction is derived in Appendix A).

We take the length of the pendulum to be $L =0.5$ m, as carbon nanotubes of this length have been demonstrated in the experimental literature \cite{zhang}.  Using this, the period of the pendulum $T$ is on the order of 1.42 s.  We will also assume that the preparation and recombination of the spatial superposition occur on timescales which are negligible compared to the time $\tau$ over which the gravitational interaction is considered.  The relevant condition for the short-time approximation is therefore $\tau/T\ll1$.  The interaction time $\tau$ must be chosen to balance maximal accumulation of gravitational phase against the requirement that the constrained dynamics remain close to their locally inertial limit.

We note that the large pendulum length also allows the required spatial
superposition sizes to be achieved within the small-angle regime.  For a
symmetric superposition with branch separation
$\Delta x\sim1\,\mu\mathrm{m}$, the two branches are displaced from the
central position by $\Delta x/2$.  The corresponding angular range is on the order of 

\begin{equation}
    \theta\sim\frac{\Delta x}{2L}\sim10^{-6},
\end{equation}

\noindent
which is well within the regime where the
small-angle approximation is valid.

Finally, the inclusion of a paramagnetic compensator can be used to reduce the effective magnetic susceptibility of the composite particle and suppress residual magnetic forces, including those responsible for diamagnetic oscillations. This improves the validity of the locally inertial approximation whilst retaining the mechanical stability provided by the pendulum geometry.  An alternative implementation without such compensation is also possible, in which case shorter nanotubes (with periods on the order of $0.2\,\mathrm{s}$) may be used in conjunction with motional dynamical decoupling techniques \cite{wood}.

%Although generation of macroscopic superpositions of a diamond has been discussed previously in the literature for free fall, there are serious problems with this type of experiment.  Firstly, it is estimated that one would need a drop time of at least $1$ s in order to observe generation of a superposition due to quantum effects, even without a pre-drop to increase the speed of the diamond. 

$\textit{Interferometry}$ - Having discussed the constrained motion, we next
consider a Stern-Gerlach-type inteferometer.  A magnetic field gradient couples the spin of the NV center to the center-of-mass motion, producing a spin-dependent force and hence a spatial splitting of the wavefunction.  To probe gravity-induced entanglement, we consider two such interferometers adjacent to each other (shown schematically in Fig. 2) \cite{bose, vedral}.  More specifically, we consider a linear two-qubit geometry where the relevant branch displacements of the two interferometers lie along a common axis (the branch coordinates are treated as signed scalar displacements along this axis) \cite{schut3}.  In contrast to free-fall implementations, the absence of diamagnetic oscillations in the present scheme allows sufficiently strong gradients without inducing additional motional instabilities.

Each system evolves in a spatial superposition and the resulting branch-dependent gravitational interaction gives rise to an entangling phase during the interferometer sequence.  For a given pair of branches $i,j \in \{L,R\}$, the interaction energy may be written as
\begin{equation}
V_{ij}(t) = -\frac{G m^2}{r_{ij}(t)},
\end{equation}
where $r_{ij}(t)$ denotes the separation along the corresponding trajectories. The resulting phase difference accumulated over the interferometer sequence is
\begin{equation}
\Delta \phi = \frac{1}{\hbar} \int_0^\tau dt \left[ V_{LR}(t) - V_{LL}(t) \right].
\end{equation}

% The Stern–Gerlach interaction produces a spin-dependent displacement of each centre of mass along the interferometric axis. We denote the two branches by $L$ and $R$, corresponding to displacements of equal magnitude and opposite sign about the equilibrium position. 

\begin{figure}
  \centering
   \includegraphics[width=85mm]{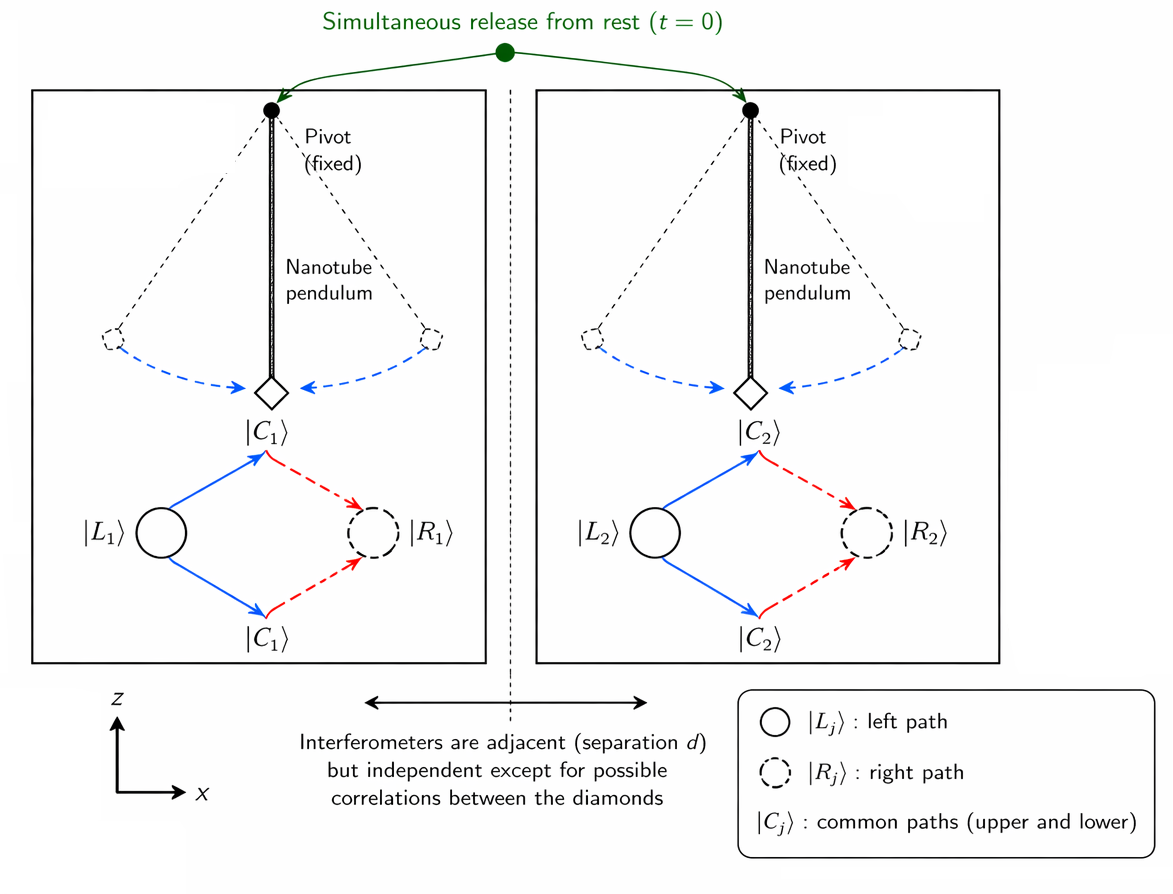}

 \caption{
Schematic of two adjacent nanotube-based interferometers for probing gravity-induced entanglement. Each interferometer contains a nanodiamond attached to a carbon nanotube pendulum, prepared in an identical initial state and released simultaneously from rest at $t=0$. A spin-dependent force generates spatial superpositions corresponding to left and right paths $|L_j\rangle$ and $|R_j\rangle$, with intermediate common paths $|C_j\rangle$. The two systems are separated by a distance $d$.  In the idealised model, the two systems are treated as independently suspended and interacting only via their mutural gravitational interaction: suppression of residual mechanical and electromagnetic cross-couplings is an experimental requirement.  As in \cite{ bose}, different branch configurations experience different gravitational potentials, leading to phase accumulation and entanglement between the two particles upon recombination.
}
  
\end{figure}

\textit{Phase correction} - The leading correction to the gravitationally induced entangling phase arises from the short-time deviation of the spin-dependent center-of-mass trajectory from uniform acceleration.  The relative correction to the trajectory $x$ is found to be

\begin{equation}
\bigg|\frac{\delta x}{x_{\mathrm{free}}} \bigg|
\sim \frac{\pi^2}{3}\frac{t^2}{T^2},
\end{equation}
where $x_{\text{free}} = (1/2) a t^2$ is the corresponding uniformly accelerated trajectory.  For completeness, the correction is derived in Appendix A.

The constrained trajectory modifies the entangling phase because the branch-dependent separations entering the interaction integral are altered.  For each branch separation, we write

\begin{equation}
  r_{ij} (t) = r_{ij}^{\text{free}} (t) + \delta r_{ij} (t)  ,
\end{equation}

\noindent
where $ r_{ij}^{\text{free}} (t)$ denotes the uniformly accelerated case and $ \delta r_{ij} (t)$ denotes the corrected version using the trajectory derived in Appendix A.  Expanding the gravitational interaction to first order gives a correction to each branch phase

\begin{equation}
  \delta \phi_{ij} = \frac{Gm^2}{\hbar} \int_0^{\tau} d t \: \frac{\delta r_{ij} (t)}{r^2_{ij}(t)}  . 
\end{equation}

\noindent
The gravitationally induced entanglement is encoded into the relative Bose-Marletto-Vedral phase
$\Delta\phi = \phi_{LL} + \phi_{RR} - \phi_{LR} - \phi_{RL}$, which determines the final spin state after recombination.  The correction to this phase is evaluated explicitly in Appendix B and found to be
 
\begin{equation}
  \frac{\delta (\Delta \phi)}{\Delta \phi_{\text{free}}} \sim C \frac{\pi^2}{9} \frac{\tau^2}{T^2} , 
\end{equation}

\noindent
where $C$ is a geometric factor determined by the specific interferometer configuration.  If we assume a symmetric configuration, the dependence of $C$ on the ratio of the branch displacement to the equilibrium separation gives $C \sim O(1)$, hence the correction is controlled primarily by the short-time parameter $\tau^2/T^2$.

The corresponding two-qubit state may be written up to local phases as
\begin{equation}
|\psi\rangle \propto \sum_{i,j \in \{L,R\}} e^{i\phi_{ij}} |ij\rangle.
\end{equation}
\noindent

\noindent
We assume that all non-gravitational sources of noise (thermal motion, spin dephasing, and mechanical fluctuations) can be absorbed into an effective visibility decay $V \to V e^{-\Gamma t}$, with $\Gamma$ capturing the dominant environmental decoherence channels. Since the gravitational phase is unaffected by these processes to leading order, decoherence enters only as an overall reduction in contrast and does not modify the accumulation of entanglement phase.

$\textit{Timescale compatibility and magnitude of phase}$ - We first distinguish the timescale associated with preparation of the spatial superposition from the subsequent gravitational interaction time $\tau$.  The free-flight interferometeric protocol of Wan et al. provides an example of a spin-dependent scheme for which the spatial superposition can be prepared and recombined on sub-millisecond timescales \cite{wan}.  More generally, rapid preparation of spatial superpositions has been considered in a range of proposals and it is considered feasible to realise superpositions with masses between $10^{-17}$ and $10^{-14}$ kg and superposition sizes on the order of micrometres \cite{xiang}.

The magnitude of the gravitational phase must be considered separately.  From Appendix B, the BMV phase in the symmetric configuration is

\begin{equation}
    \Delta \phi_{\text{free}} \sim - \frac{Gm^2 \tau}{\hbar} \bigg( \frac{2}{d} - \frac{1}{d+ \Delta x} - \frac{1}{d - \Delta x}\bigg)  .
\end{equation}

\noindent
As a benchmark, the parameters quoted by Wan et al. are $m \sim 1.25 \times 10^{-17}$ kg and $\Delta x \sim 100$ nm.  Combined with representative values $d =450 \mu$m and $\tau = 0.1$ s, we find that $\Delta \phi_{\text{free}} \sim 10^{-15}$ rad \cite{wan}.  This demonstrates that the feasibility of preparing and recombining a spatial superposition in their protocol should not be conflated with the feasibility of generating an appreciable entangling phase through the gravitational interaction of two test masses. 

We therefore consider a separate regime motivated for rapid generation of macroscopic spatial superpositions, taking $m \sim 10^{-14}$ kg, $\Delta x \sim 1$ $\mu$m, and $d \sim 100$ $\mu$m.  For these parameters and taking as a representative interaction time $\tau = 0.14 $s, equation (14) evaluates to $\Delta \phi_{\text{free}} \sim 10^{-3}$ rad.  Using the correction derived in Appendix B, we find that $\delta ( \Delta \phi) / \Delta \phi_{\text{free}} \sim 10^{-2}$ and the corresponding absolute correction $\delta (\Delta \phi)$ is therefore on the order of $10^{-5}$ rad, hence the constraint-induced modification is negligible compared with the entangling phase itself.  

Although the BMV phase is small, this does not imply that the effect is inaccessible to measurement.  Signal amplification protocols have been proposed specifically to enhance the observability of gravity-induced entanglement, where a small gravitational phase is boosted to achieve a measurable signal \cite{streltsov, feng}.  In the present case, an enhancement factor of $10^3$ to the effective gravity-induced coupling strength (considered  feasible in \cite{streltsov}) would convert the phase to order unity whilst leaving the constraint-induced correction unchanged, providing a potential route to an experimentally accessible signal.  Feng and Vedral also observe that their readout amplification strategy can be used to recover observable entanglement even when the interaction time is reduced by one or two orders of magnitude, the exact scenario which we describe in this work \cite{feng}.  The feasibility of the underlying interferometric dynamics is therefore established independently of the amplification scheme, which is relevant to the subsequent task of extracting the small gravitational phase experimentally.

\textit{Non-gravitational interactions and mechanical isolation} - The analysis above assumes that the two suspended test masses evolve independently except for their mutual gravitational interaction.  This is a non-trivial assumption which is important for interpreting the generated entanglement as a
signature of gravity and requires suppression of additional interaction channels.  The local constraint force associated with each nanotube does not by itself provide an interaction between the two test masses.  

We assume that the two nanotubes are mounted on mechanically independent supports, with no shared support degree of freedom or other mechanical coupling between the two suspension systems.  Under this assumption, the constraint Hamiltonian is local with respect to the bipartition of the two masses:

\begin{equation}
  H_{\text{constr}} =  H_{\text{constr},1} \otimes I_2 +    H_{\text{constr},2} \otimes I_1 .
\end{equation}

\noindent  
The corresponding evolution therefore factorizes into local unitaries and cannot generate entanglement from an initially separable state. This assumption is important in BMV-type experiments, where uncontrolled couplings and acceleration noise can constitute unwanted background for the gravity-mediated entanglement signal \cite{krisnanda}.  Any entanglement generated by the constraint/support sector would instead require a nonlocal coupling between the two suspension systems.  Even if the two nanotubes are mounted on independent supports, there could still be residual mechanical cross-coupling due to vibrations which we require to be below the gravitational interaction scale.  A quantitative study of the mechanical noise in this case is beyond the scope of this work, but this requirement is analogous to ones identified in the literature on noise mitigation in experimental QGEM protocols \cite{toros}.

Electromagnetic interactions provide a second class of potential
non-gravitational channels.  These include electrostatic forces, magnetic
dipole interactions, induced electromagnetic interactions involving the nanotubes and the paramagnetic compensators, and couplings
associated with the magnetic-field gradients used for spin-dependent
splitting.  Such effects are not included in the idealized Hamiltonian
considered here and must be independently suppressed or characterized.
In particular, electromagnetic screening, magnetic
shielding, and appropriate choice of spin manipulation protocols are
standard requirements in proposed QGEM experiments and have been studied and characterised extensively for various geometries, including the linear two-qubit setup used in this work
\cite{schut2, schut3}.  

We emphasise that the purpose of the present work is to determine how a mechanical constraint
modifies gravity-induced entanglement once the two systems are
treated as independent.  We therefore regard the suppression of
cross-coupling between the mechanical suspensions and electromagnetic backgrounds as
an experimental requirement and not as an additional dynamical
interaction in the model, which is beyond the scope of this work.  Importantly, the existing literature on these effects shows that they can in principle be
distinguished from the gravitational contribution through their different
dependence on separation, shielding, and experimental control parameters \cite{schut4}.

$\textit{Internal vibrations}$ - As mentioned above, a key problem with the free fall scheme is the large drop height required.  In contrast to a free fall experiment where the relevant length scale is set by the full drop height, the nanotube pendulum operates in a regime where the trajectory is defined by a fixed mechanical structure which can be equilibrated and left undisturbed over long timescales.  Thermal expansion is in principle still present for the nanotube, but it only leads to a static renormalization of the nanotube length once thermal equilibrium is reached, instead of introducing run-to-run fluctuations.

It is important to note that the constrained platform described in this article possesses two distinct types of mechanical degrees of freedom, both of which can be distinct sources of noise which interfere with observation of entanglement.  The first is the low frequency rigid pendulum mode, which determines the center-of-mass trajectory entering into the interferometer dynamics. The second consists of internal elastic modes of the nanotube (flexural, torsional and longitudinal), which describe small deformations about the rigid trajectory.  

We will deal with the internal modes first.  The fundamental flexural mode may be modelled as a harmonic oscillator with mechanical frequency $\omega_m$ and effective mass $m_{\mathrm{eff}}$. In thermal equilibrium, the mean-square displacement is given by
\begin{equation}
\langle u^2 \rangle = \frac{k_B T}{m_{\mathrm{eff}} \omega_m^2},
\end{equation}
where $k_B$ is the Boltzmann constant, $ T$ is the temperature and $u$ is the transverse displacement of the nanotube due to elastic deformation.  Carbon nanotubes typically exhibit mechanical frequencies in the range of kHz to MHz, and the system may be operated at cryogenic temperatures on the order of 10 mK \cite{cao1, cao2, wu, zhao}.

%  leading to extremely small displacement on the length scales relevant for the interferometer dynamics 

These fluctuations occur on timescales set by $1/\omega_m$, which are much shorter than the interferometer evolution time, so their contribution to the phase noise is strongly suppressed.  To estimate the magnitude of the fluctuations, we take representative parameter values $T = 10^{-2}$ K, $\omega_m = 10^5$ $\mathrm{s}^{-1}$, and $m_{\mathrm{eff}} = 10^{-14}$ kg. This gives
\begin{equation}
\sqrt{\langle u^2 \rangle}
\sim \sqrt{\frac{k_B T}{m_{\mathrm{eff}} \omega_m^2}}
\sim 10^{-11}\,\mathrm{m},
\end{equation}
corresponding to fluctuations on the order of picometers, which are negligible compared to the spatial superposition scale.  Higher flexural modes of the nanotube have larger eigenfrequencies \cite{treacy}.  Since the thermal amplitude of each mode scales approximately as $1 / \omega_n$, their contribution to the displacement of the particle is smaller than that of the fundamental mode, which provides the dominant contribution in the equilibrium regime.

In the linear beam approximation, the torsional and longitudinal modes do not contribute to transverse displacement.  This approximation may not be valid for the extremely slender nanotubes considered here, since nonlinearities can couple the modes when transverse deflections of the beam become appreciable.  However, analysis of nonlinear beam dynamics shows that the induced transverse displacement due to torsional and longitudinal modes is smaller than the displacement associated with the fundamental flexural mode in the parameter regime considered here \cite{carbonara}.  We therefore conclude that thermal vibrations of the nanotube do not constitute a significant source of noise in the proposed experiment.

\textit{Thermal motion of the pendulum coordinate} - The above estimate only deals with the internal mechanical degrees of freedom of the nanotube but we must also consider the pendulum coordinate itself.  This defines the slowly varying equilibrium position, whereas the interferometer trajectory is determined by the spin-dependent displacement about this equilibrium.  We therefore consider its contribution to the gravitational phase separately.  We begin by writing the phase

\begin{equation}
 \phi = -\frac{1}{\hbar} \int_0^{\tau} d t \: \frac{Gm^2}{r(t)}   
\end{equation}

\noindent
and assume that the separation $r$ along the trajectories is perturbed by $\delta  r^{\text{th}}(t)$, where $\delta r^{\text{th}} \ll r$.  The perturbed phase is   

\begin{equation}
 \phi = -\frac{1}{\hbar} \int_0^{\tau} d t \: \frac{Gm^2}{r(t) + \delta r^{\text{th}}(t)} .  
\end{equation}

\noindent
Taylor expanding the reciprocal by first-order perturbation theory and substituting back in, we find that 

\begin{equation}
  \delta \phi^{\text{th}}  = \frac{Gm^2}{\hbar} \int_0^{\tau} d t \: \frac{\delta r^{\text{th}} (t)}{r(t)^2}  . 
\end{equation}

We next assume that the branch separation only varies weakly during the relevant part of the interferometric trajectory so that $r(t) \approx d$, where $d$ is the characteristic separation between the masses.  We then assume that the perturbation is approximately constant for each experimental run, so that $  \delta r^{\text{th}}(t) \approx \delta d^{\text{th}}$.  From this, we obtain 

\begin{equation}
  \delta \phi^{\text{th}} \approx \frac{Gm^2 \tau}{\hbar d^2} \delta d^{\text{th}},  
\end{equation}

\noindent
from which we may obtain an expression for the phase uncertainty

\begin{equation}
 \bigg| \frac{\delta \phi^{\text{th}}}{\phi}  \bigg|   \approx \frac{\delta d^{\text{th}}}{d}.
\end{equation}

For representative values, $d$ can be taken on the order of 100 $\mu$m and the fluctuation of the separation $d$ which arises due to the thermal motion of the center of mass is taken to be on the order of 1 $\mu$m, implying that $\delta \phi^{\text{th}} / \phi$ is of order $10^{-2}$.  This matches the value found in \cite{bose} for the change in the gravitational phase due to center-of-mass thermal motion from the warmth of the trap, hence in this respect our protocol is consistent with that of Bose-Marletto-Vedral and does not imply a larger modification to the gravitational phase than the one they found when using a relatively high initial temperature and release from a 1 MHz trap. 

Finally, we note that the thermal displacement of the pendulum should be interpreted here as a slowly varying initial condition, since the center-of-mass displacement changes negligibly during a single experimental run.  The contribution from thermal motion therefore appears primarily as a run-to-run variation in the initial separation between the interferometers which may be calibrated for or averaged over repeated measurements and does not constitute a significant source of dynamical phase noise over one interferometric sequence.

% Since the interferometer sequence is completed on a timescale $t \sim 10^{-4}-10^{-3}$ s whilst the period of the pendulum is $T \sim 1$ s,

%Thermal fluctuations of this low-frequency mode do not represent rapid noise during a single interferometric sequence, but rather correspond to an uncertainty in the initial conditions of the trajectory. For a sequence of duration $t<<T$, the resulting uncertainty enters as a slowly varying offset of displacement and velocity and not as an oscillatory source of phase noise. 

\textit{Visibility correction} - Entanglement detection is based on the interference visibility $V=\cos(\Delta\phi)$, with entanglement certified when $V$ exceeds the separability bound \cite{bose, vedral}.  The effect of the constrained trajectory on the visibility follows from the corresponding correction to the gravitational phase.  Writing

\begin{equation}
\Delta \phi_{\text{pend}} = \Delta \phi_{\text{free}} + \delta ( \Delta \phi),    
\end{equation}

\noindent
the visibility becomes 

\begin{equation}
V_{\mathrm{pend}} = \cos\!\left(\Delta\phi_{\mathrm{free}} + \delta ( \Delta \phi) \right).
\end{equation}

\noindent
For $|\delta (\Delta \phi)| \ll 1 $, expansion to first order gives

\begin{equation}
V_{\mathrm{pend}} = V_{\mathrm{free}}
- \sin(\Delta\phi_{\mathrm{free}})\,\delta (\Delta \phi)
+ \mathcal{O}((\delta ( \Delta \phi))^2),
\end{equation}

\noindent
where $V_{\text{free}} = \cos ( \Delta \phi_{\text{free}})$.  Since $|\sin(\Delta\phi_{\mathrm{free}})| \le 1$, the visibility correction satisfies

\begin{equation}
|V_{\mathrm{pend}} - V_{\mathrm{free}}|  \lesssim | \delta ( \Delta \phi)|.
\end{equation}

\noindent
For the representative parameters considered above, $| \delta ( \Delta \phi)| \sim 10^{-5}$ rad and therefore the visibility correction is bounded at the level of $10^{-5}$.  This bound is independent of whether the gravitational phase is subsequently enhanced using an amplification protocol.

$\textit{Conclusions}$ - To summarise, we have shown that a mechanically constrained system can reproduce the dynamical conditions required for free-fall interferometry, thereby realising a locally inertial regime within a bound configuration. In this regime, the center-of-mass motion is constrained whilst conserving the gravitational phase accumulation central to proposals for witnessing the quantum nature of gravity \cite{bose, vedral}.  Crucially, however, the constraint-induced corrections enter via the small parameter $(t/T)^2$ and do not significantly modify the operational entanglement witness in experimentally relevant regimes.  

As a result, the present implementation replaces the dominant sources of noise in free-fall experiments (uncontrolled macroscopic drift and run-to-run trajectory variations) with bounded fluctuations which can be systematically averaged over long integration times.  This shift in the underlying noise structure and presence of a well-defined, reproducible phase accumulation enables stable interferometry.  The proposal therefore identifies a physically motivated constrained regime where the uniform acceleration approximation remains accurate, providing a platform which is complementary to existing approaches for enhancing the observability of gravity-induced entanglement.

As mentioned earlier, an interesting and important direction for future work would be to combine the constrained platform proposed here with recent schemes designed to enhance the observability of gravity-induced entanglement.  Since the present work addresses one of the main experimental obstacles for entanglement by replacing free fall with a constrained trajectory, it is complementary to proposals which enhance the signal using massive mediators or entanglement amplification protocols based on weak values \cite{streltsov, feng}.  Since the interaction time is short in the proposed experiment and the gravitational phase is correspondingly small, it will likely be necessary to use these techniques to extract an observable signal.  Investigating whether such techniques can be incorporated into these platforms may provide a route towards experimentally accessible tests of gravity-induced entanglement.

%We also emphasise that in the pendulum setup levitation and trapping of micron-sized particles is no longer required.

\textit{Acknowledgments -} HW thanks Gavin Morley for suggesting the initial idea for this problem and Jonte Hance for useful discussions.  This research was partly conducted whilst HW was visiting the Okinawa Institute of Science and 
Technology (OIST) through the Theoretical Sciences Visiting Program (TSVP).  HW acknowledges support from a London Mathematical Society Early Career Research Travel Grant (ECR-2526-51).

\textit{Appendix A: Deviation from uniformly accelerated trajectory} - Consider the spin-dependent center-of-mass displacement $x(t)$.  In the mechanically constrained system, this obeys

\[   \ddot{x} + \omega^2 x = a ,     \tag{A1}\]

\noindent
where $a$ is the effective Stern-Gerlach acceleration and $\omega = 2 \pi /T$.  Using as initial conditions $x(0)= 0$ and $\dot{x} (0) = 0$, the solution is

\[ x(t) = \frac{a}{\omega^2}(1 -\cos \omega t) .    \tag{A2}\]

%\[ \theta (t) = \theta_0 \cos (\omega t) .    \tag{A2}\]

\noindent
The short-time expansion of equation (A2) is

\[  x(t) = \frac{1}{2} a t^2 - \frac{a \omega^2}{24} t^4 + O (t^6).    \tag{A3} \]

%\[  \theta (t) = \theta_0 \bigg( 1 - \frac{\omega^2 t^2}{2} + \frac{\omega^4 t^4}{24} + ... \bigg)     \tag{A3} \]

\noindent
The uniformly accelerated trajectory is given by 

\[   x_{\text{free}} = \frac{1}{2} at^2  .        \tag{A4}\]

\noindent
Taking the difference, we therefore have 

\[  \frac{\delta x}{x_{\text{free}}}  = - \frac{\omega^2 t^2}{12} = - \frac{\pi^2}{3} \frac{t^2}{T^2} .   \tag{A5}\]

\textit{Appendix B: Correction to the gravitational phase} -  We start by writing down the branch phase:

\[    \phi_{ij} = - \frac{Gm^2}{\hbar} \int^{\tau}_0 \frac{dt}{r_{ij}(t)}  .     \tag{B1}  \]

\noindent
From the constrained trajectory, we have

\[  r_{ij} (t) = r_{ij}^{\text{free}} (t) + \delta r_{ij}(t),       \tag{B2}\]

\noindent
where

\[ \delta r_{ij}(t) = \delta x_j (t) - \delta x_i(t)  .        \tag{B3}\]

\noindent
$x_i$ and $x_j$ are signed displacements along the interferometer axis.  In Appendix A, we observed that

\[  \delta x_i (t) = - \epsilon x_i^{\text{free}} (t)          \tag{B4} \]

\noindent
with

\[    \epsilon(t) = \frac{\pi^2}{3} \frac{t^2}{T^2} .     \tag{B5}  \]

Substituting into equation (B3), we obtain

\[  \delta r_{ij}(t)  = - \epsilon(t) ( x_j^{\text{free}} (t)  -x_i^{\text{free}} (t) ).        \tag{B6}  \]

\noindent
We take the two interferometers to be arranged in the linear two-qubit configuration, with their equilibrium centres separated by a distance $d$ along a common interferometric axis \cite{schut3}.  The branch labels $i,j \in \{L,R \}$ denote signed displacements along this axis.  Using

\[  r_{ij}^{\text{free}}  (t)  = d + x_j^{\text{free}} (t)   -x_i^{\text{free}} (t)  ,    \tag{B7}   \]

\noindent
where $d$ is the equilibrium separation, we arrive at 

\[    \delta r_{ij}(t) = - \epsilon (t) (  r_{ij}^{\text{free}}  (t)  - d)   .  \tag{B8}\]

\noindent
After expanding the phase, the perturbation is found to be

\[      \delta \phi_{ij}= \frac{Gm^2}{\hbar} \int^{\tau}_0 dt \: \frac{\delta r_{ij}(t)}{(r_{ij}^{\text{free}}(t))^2}       . \tag{B9}  \]

\noindent
Substituting in equation (B8), we finally get

\[      \delta \phi_{ij}= -\frac{Gm^2}{\hbar} \int^{\tau}_0 dt \:  \epsilon(t) \frac{r_{ij}^{\text{free}} (t) - d}{(r_{ij}^{\text{free}}(t))^2}       . \tag{B10}  \]

To detect entanglement, we must consider the full Bose-Marletto-Vedral phase $\Delta \phi_{}$:

\[    \Delta \phi = \phi_{LL} + \phi_{RR} - \phi_{LR} - \phi_{RL}  .    \tag{B11}\]

\noindent
The perturbation to the phase is therefore

\[        \delta ( \Delta \phi) =  -\frac{Gm^2}{\hbar} \int^{\tau}_0 dt \:  \epsilon(t) \bigg[   \frac{r_{LL}^{\text{free}} (t) - d}{(r_{LL}^{\text{free}}(t))^2}  +  \frac{r_{RR}^{\text{free}} (t) - d}{(r_{RR}^{\text{free}}(t))^2} \]

\[ -  \frac{r_{LR}^{\text{free}} (t) - d}{(r_{LR}^{\text{free}}(t))^2}  -  \frac{r_{RL}^{\text{free}} (t) - d}{(r_{RL}^{\text{free}}(t))^2}       \bigg]      . \tag{B12} \]

\noindent
From equation (B1), the corresponding unperturbed phase is 

\[  \Delta \phi_{\text{free}} =   - \frac{Gm^2}{\hbar} \int^{\tau}_0  dt \: \bigg[         \frac{1}{r_{LL}^{\text{free}}(t)}  + \frac{1}{r_{RR}^{\text{free}}(t)} - \frac{1}{r_{LR}^{\text{free}}(t)}   -\frac{1}{r_{RL}^{\text{free}}(t)}      \bigg]   .  \tag{B13}\]

\noindent
The relative correction to the phase is therefore found to be

\begin{widetext}

\[   \frac{ \delta ( \Delta \phi)}{ \Delta \phi_{\text{free}}}      =  \frac{\int^{\tau}_0 dt \:  \epsilon(t) \bigg[   \frac{r_{LL}^{\text{free}} (t) - d}{(r_{LL}^{\text{free}}(t))^2}  +  \frac{r_{RR}^{\text{free}} (t) - d}{(r_{RR}^{\text{free}}(t))^2} -  \frac{r_{LR}^{\text{free}} (t) - d}{(r_{LR}^{\text{free}}(t))^2}  -  \frac{r_{RL}^{\text{free}} (t) - d}{(r_{RL}^{\text{free}}(t))^2}       \bigg] }{ \int^{\tau}_0  dt \: \bigg[         \frac{1}{r_{LL}^{\text{free}}(t)}  + \frac{1}{r_{RR}^{\text{free}}(t)} - \frac{1}{r_{LR}^{\text{free}}(t)}   -\frac{1}{r_{RL}^{\text{free}}(t)}      \bigg] } .  \tag{B14}  \]

\end{widetext}

To simplify this general expression, we may assume for the uniformly accelerated trajectory that the change in the branch separation during the phase accumulation time is very small and that the leading correction can be evaluated using a characteristic separation $ r_{ij}^{\text{free}}$, where $ r_{ij}^{\text{free}}(t) \sim r_{ij}^{\text{free}}$. This enables us to evaluate the integrals and we arrive at

\[   \frac{ \delta ( \Delta \phi)}{ \Delta \phi_{\text{free}}} \sim \frac{\pi^2}{9} \frac{\tau^2}{T^2}  \frac{\frac{r_{LL}^{\text{free}} - d}{(r_{LL}^{\text{free}})^2}  + \frac{r_{RR}^{\text{free}} - d}{(r_{RR}^{\text{free}})^2}  - \frac{r_{LR}^{\text{free}} - d}{(r_{LR}^{\text{free}})^2}   - \frac{r_{RL}^{\text{free}} - d}{(r_{RL}^{\text{free}})^2}     }{\frac{1}{r_{LL}^{\text{free}}}+ \frac{1}{r_{RR}^{\text{free}}} - \frac{1}{r_{LR}^{\text{free}}} -\frac{1}{r_{RL}^{\text{free}}}}  . \tag{B15}  \]

%
%To evaluate the geometry-dependent coefficient, it is necessary to specify a configuration for the experiment.  For a representative symmetric configuration, 
\noindent
Without loss of generality, we may take the spin-dependent displacements of each pendulum to have equal magnitude and opposite sign about their equilibrium positions.  Denoting the superposition scale by $\Delta x$, we write $x_L^{\text{free}} = -\Delta x/2$ and $x_R^{\text{free}} =\Delta x/2$.  Using equation (B7), the four branch separations are therefore $r_{LL}^{\text{free}} =r_{RR}^{\text{free}} =d$, $r_{LR}^{\text{free}} = d+ \Delta x$, and $r_{RL}^{\text{free}} = d - \Delta x$, where it is assumed that $d>\Delta x$.  Using these values, the coefficient becomes

\[   \frac{\frac{r_{LL}^{\text{free}} - d}{(r_{LL}^{\text{free}})^2}  + \frac{r_{RR}^{\text{free}} - d}{(r_{RR}^{\text{free}})^2}  - \frac{r_{LR}^{\text{free}} - d}{(r_{LR}^{\text{free}})^2}   - \frac{r_{RL}^{\text{free}} - d}{(r_{RL}^{\text{free}})^2}     }{\frac{1}{r_{LL}^{\text{free}}}+ \frac{1}{r_{RR}^{\text{free}}} - \frac{1}{r_{LR}^{\text{free}}} -\frac{1}{r_{RL}^{\text{free}}}}     = \frac{2d^2}{(\Delta x)^2 - d^2}   .          \tag{B16}  \]

\noindent
In the regime where $\Delta x\ll d$, this expression goes to $-2$ and hence is of order unity.  The relevant suppression is therefore governed by the small parameter $\tau^2 / T^2$. 

%We note that this regime is compatible with the spatial superpositions discussed in \cite{bose}, including micron-scale superpositions achievable with relatively modest magnetic field gradients.  

%  Denoting the magnitude of this displacement by $s$, we write $x_L^{\text{free}} = -s$ and $x_R^{\text{free}} = s$.  Using equation (B7), the four branch separations are therefore $r_{LL}^{\text{free}} =r_{RR}^{\text{free}} =d$, $r_{LR}^{\text{free}} = d+ 2s$, and $r_{RL}^{\text{free}} = d - 2s$, where it is assumed that $d>2s$.     

% \[   \frac{\frac{r_{LL}^{\text{free}} - d}{(r_{LL}^{\text{free}})^2}  + \frac{r_{RR}^{\text{free}} - d}{(r_{RR}^{\text{free}})^2}  - \frac{r_{LR}^{\text{free}} - d}{(r_{LR}^{\text{free}})^2}   - \frac{r_{RL}^{\text{free}} - d}{(r_{RL}^{\text{free}})^2}     }{\frac{1}{r_{LL}^{\text{free}}}+ \frac{1}{r_{RR}^{\text{free}}} - \frac{1}{r_{LR}^{\text{free}}} -\frac{1}{r_{RL}^{\text{free}}}}     = \frac{2d^2}{4s^2 - d^2}   .          \tag{B16}  \]

\end{document}